\documentclass[11pt,a4paper,notitlepage]{article}

\usepackage[margin=1in]{geometry}
\usepackage{graphicx}
\usepackage{bm}
\usepackage{amssymb}
\usepackage{amsmath}
\usepackage{physics}
\usepackage{xspace}
\usepackage{titlesec}
\usepackage{xfp}
\usepackage{cite}
\usepackage{url}
\usepackage{authblk}
\usepackage{hyperref}

\newcommand{\R}{\mathbb{R}}
\newcommand{\bvec}[1]{\mathbf{#1}}

\title{
{\LARGE\bfseries
From Pixels to Light Rays: Exploring Matrix Operations
and Transformations Through Interactive Visualizations}
\\[0.3cm]
{\large\normalfont\textbf{\textit{
MathLens: An Educational Application
for Learning Linear Algebra}}}
}

\author[1]{Krishana Yadav}
\author[1]{Naman Kumar}
\author[1]{Mayank Soni}
\author[2]{Payal Wadhwa\thanks{Corresponding author:
\texttt{payal\_wadhwa@iiitvadodara.ac.in}}}

\affil[1]{Department of Computer Science and Engineering,
Indian Institute of Information Technology Vadodara, Gandhinagar - 382028,
Gujarat, India}

\affil[2]{Department of Applied Physics,
Indian Institute of Information Technology Vadodara, Gandhinagar - 382028,
Gujarat, India}

\date{}

\begin{document}

\maketitle

\vspace{-3.5em}

\begin{abstract}
 In this article, we introduce \textit{MathLens: Interactive Mathematics} \& \textit{Linear Algebra Lab}, an educational application designed to explore matrices and their transformations through digital images and extend them to elementary ray optics.\ Beginning with grayscale and color pixel representations, the application gradually explores the effect of geometric transformations and various matrix operations, including addition, subtraction, multiplication, determinants, and inverses, on image pixels, which is further extended to the reflection of light rays by plane mirrors in ray optics. By connecting numerical representations, geometric transformations, and physical phenomena within a single progressive sequence, \textit{MathLens} offers a visually motivated approach that can complement conventional learning of linear algebra.
\end{abstract}

\section{Introduction}
Matrices are among the most widely used tools in physics \cite{boas2005mathematical}, engineering \cite{kreyszig2011advanced}, computer graphics \cite{hughes2014graphics,golub2013matrix,marschner2021graphics}, and scientific computing \cite{strang2016,boyd2018applied,lay2016}. However, they are often introduced as arrays of numbers and algebraic operations, with limited emphasis on their geometrical interpretations and practical applications.\ Consequently, students may learn to perform matrix operations without fully understanding their underlying significance. 

Digital image processing provides an effective way to connect these mathematical concepts with visual experiences, as every digital image is fundamentally a structured collection of numerical values.\ A grayscale image can be represented as a two-dimensional array of pixel intensities, while a color image can be described using three separate arrays corresponding to its red, green, and blue channels \cite{gonzalez2018image}.\ Applying mathematical operations to these numerical representations can produce visible changes in brightness, contrast, orientation, and image geometry, making their effects easier to visualize. This connection motivated us to develop an interactive educational application, \textit{MathLens} \cite{mathlens2026}, designed to bridge the gap between abstract matrix operations and their visual interpretations through image-based activities. This approach is further extended to elementary ray optics, illustrating how rotation and reflection matrices can be used to represent changes in the direction of light rays \cite{ghatak2025optics,pedrotti2017optics}.

The central objective of this article is to present a visual and activity-based approach to learning linear algebra, connecting mathematical expressions with their geometrical interpretations and applications in everyday digital technologies and physical phenomena.


\section{Theoretical Framework and Visualization in MathLens}
\textit{MathLens} connects mathematical concepts with their visual interpretations through three main components:
\begin{enumerate}
    \item \textbf{Mathematical Representation:} Digital images are represented as numerical matrices, while geometrical transformations are described using coordinates and transformation matrices. 
    \item \textbf{Mathematical Operations:} Operations such as matrix addition, subtraction, multiplication, scaling, rotation, reflection, determinant, inverses, etc. are performed on the corresponding matrices or coordinates.
    \item \textbf{Interactive Visualization:} The results are displayed visually, allowing learners to vary mathematical parameters and observe their effects on images and geometrical shapes in real time.
\end{enumerate}

Together, these components provide a framework for exploring the relationship between mathematical concepts and their visual outcomes.

\section{Matrix Operations and their Visual Representations}
\subsection{Grayscale Image as a Matrix}

A grayscale image of size $M \times N$ can be represented as a matrix $\mathbf{A}$, where each element $a_{ij}$ represents the intensity of the pixel at the \(i\)-th row and \(j\)-th column \cite{gonzalez2018image}. Mathematically, it can be expressed as

\begin{equation}
\mathbf{A} \in \R^{M \times N},
\qquad
a_{ij} \in [0,255] \cap \mathbb{Z}
\label{eq:grayscale-matrix}
\end{equation}

For an 8-bit grayscale image, pixel intensities range from integral values from $[0, 255]$, where
\begin{itemize}
    \item $0 = \text{black}$ (minimum brightness).
    \item $255 = \text{white}$ (maximum brightness).
    \item Values between $0$ and $255$ represent different shades of gray, with smaller values closer to $0$ appear darker and larger values closer to $255$ appear brighter.
\end{itemize}

\begin{figure}[t]
 \caption{Grayscale image and their corresponding $4 \times 4$ matrix representations: (a) a complete black image, (b) a completely white image, (c) an image with varying grayscale intensities, and (d) the effect of scalar multiplication on image intensity.}
\centering
    \includegraphics[width=1.01\columnwidth,keepaspectratio]{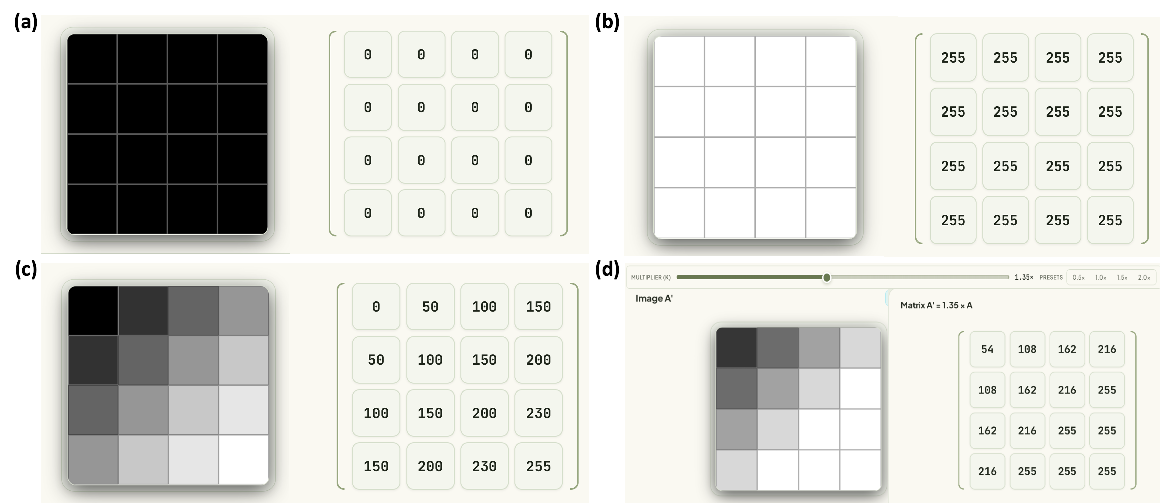}
    \label{fig:1}
\end{figure}

For example, a completely black $4 \times 4$ image can be represented by the zero matrix $\mathbf{O}_{4 \times 4} = [0]_{4 \times 4}$ and a completely white image is represented by a matrix in which every element is 255, i.e., $[255]_{4 \times 4}$, as shown in Figure \ref{fig:1}(a) and \ref{fig:1}(b), respectively.\ Similarly, different shades of gray in a grayscale image can be represented by a matrix with pixel values ranging from $0$ to $255$, as shown in Figure \ref{fig:1}(c).\ 

\subsection{Scalar Multiplication and Brightness}
\label{sec:scaling}
Scalar multiplication provides a simple way to adjust the brightness of an image.\ If we multiply a matrix $\textbf{A}$ with a scalar $k$, it changes the intensity of every pixel \cite{gonzalez2018image}, as given by
\begin{equation}
\mathbf{A}' = k\mathbf{A}
\end{equation}
where $k$ is the scaling factor.
\begin{itemize}
    \item When $k > 1$ (e.g., $k=1.5$ or $k=2.0$), all pixel intensities increase uniformly, making the overall image brighter.
    \item When $0 < k < 1$ (e.g., $k=0.5$), all pixel intensities decrease uniformly, making the overall image darker.
    \item When $k = 1$, the image remains unchanged.
\end{itemize}
The corresponding change in the image brightness of the grayscale image for the scaling factor $k=1.35$ is shown in Figure \ref{fig:1}(d).\ \textit{MathLens} provides an interactive slider to change the scaling factor $k$ and observe how scalar multiplication affects image brightness \cite{mathlens2026}.\ Since pixel intensities in an 8-bit image must lie between 0 (black) and 255 (white), multiplying them by a scaling factor may produce values outside this range. Therefore, values exceeding 255 are clipped to 255 and the resulting intensities are rounded to the nearest integer \cite{gonzalez2018image}. 

\subsection{From Grayscale to RGB Color Vectors}
Unlike a gray pixel, which is represented by a single intensity value, a color pixel is represented by three independent values corresponding to the intensities of red ($R$), green ($G$), and blue ($B$) \cite{gonzalez2018image}, and can be written as a 3$\times$1 column vector given by 

\begin{equation}
\bvec{v} = \begin{bmatrix} R \\ G \\ B \end{bmatrix} \in [0, 255]^3 \subset \R^3
\end{equation}

For an 8-bit RGB image, each component ranges from 0 to 255. Therefore, the three primary colors can be represented as 
\begin{equation}
 \begin{aligned}
 \mathbf{v}_{\text{Red}} &= \begin{bmatrix} 255 \\ 0 \\ 0 \end{bmatrix}, \quad
\mathbf{v}_{\text{Green}} = \begin{bmatrix} 0 \\ 255 \\ 0 \end{bmatrix}, \quad
\mathbf{v}_{\text{Blue}} = \begin{bmatrix} 0 \\ 0 \\ 255 \end{bmatrix}\\   
\end{aligned}   
\end{equation}

\begin{figure}[t]
 \caption{Color image and their corresponding $3 \times 1$ RGB vector  representations: (a) a completely red image and (b) a purple image.}
\centering
    \includegraphics[width=1.01\columnwidth,keepaspectratio]{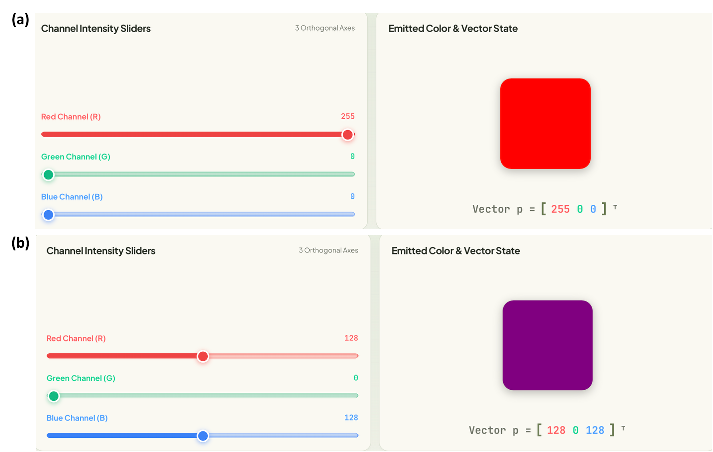}
    \label{2}
\end{figure}

Further, different colors can be obtained by relative intensity of these three components.\ For example, purple can be represented as 
\begin{equation}
 \begin{aligned}
\mathbf{v}_{\text{Purple}} &= \begin{bmatrix} 128 \\ 0 \\ 128 \end{bmatrix} \quad (\text{equal mixture of Red and Blue)}
\end{aligned}   
\end{equation}

where, red and blue components have equal intensities. The corresponding visual representations of red and purple are shown in Figure \ref{2}(a) and \ref{2}(b), respectively.\ \textit{MathLens} provides three independent interactive sliders for $R$, $G$, and $B$, allowing learners to observe how different combinations produce different colors \cite{mathlens2026}.
Extending this idea from a single pixel to an entire image of dimension $M \times N$, a color image can be represented as the sum of its red, green, and blue component matrices, given by
\begin{equation}
\begin{aligned}
\mathbf{I}=\mathbf{I}_{Red}+\mathbf{I}_{Green}+\mathbf{I}_{Blue}
\end{aligned}
\end{equation}

Here, each component matrix contains RGB vectors with only its corresponding non-zero color channel given by
\begin{equation}
\begin{aligned}
\mathbf{I}_{\text{Red}} &= (\mathbf{R}, \mathbf{0}, \mathbf{0}), \quad
\mathbf{I}_{\text{Green}} = (\mathbf{0}, \mathbf{G}, \mathbf{0}),\\
\mathbf{I}_{\text{Blue}} &= (\mathbf{0}, \mathbf{0}, \mathbf{B})
\end{aligned}
\end{equation}

\begin{figure}[t]
 \caption{(a) A $4 \times 4$ color image and its corresponding red, green, and blue (RGB) component matrices,(b) Scalar multiplication of a \(4 \times 4\) image matrix by a factor of \(k = 1.55\) and its corresponding visual effects.}
\centering
    \includegraphics[width=1.01\columnwidth,keepaspectratio]{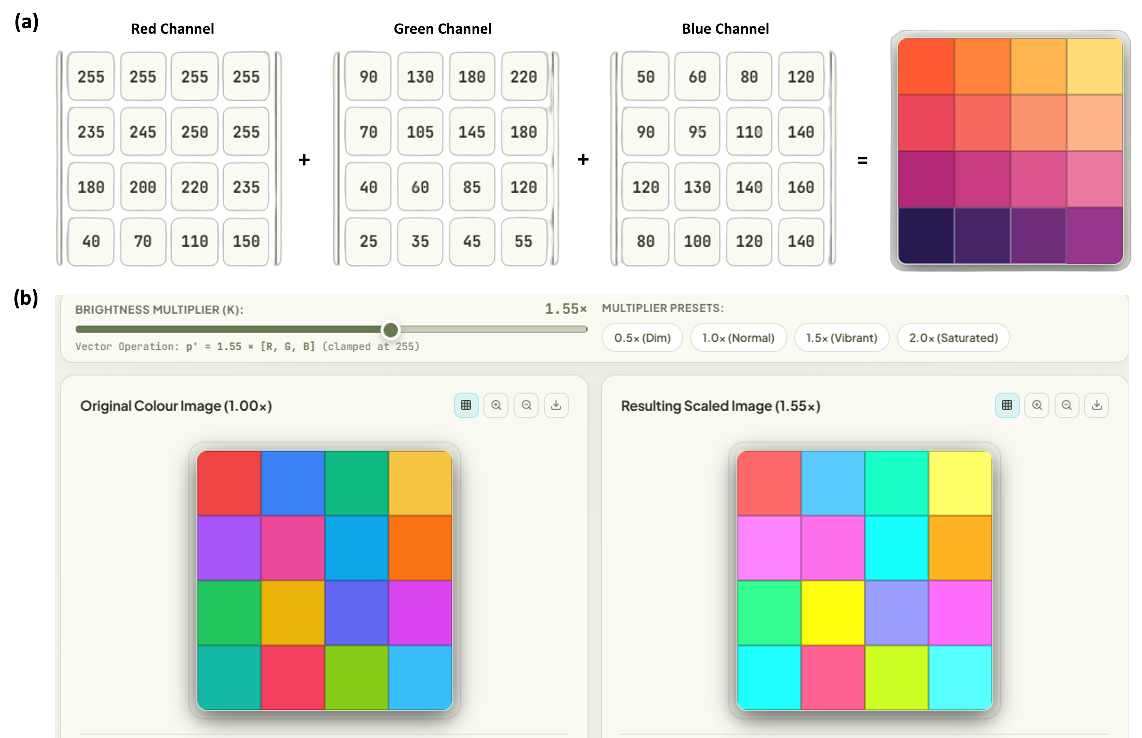}
    \label{3}
\end{figure}

\textit{MathLens} provides an interactive animation showing the color image separating into its independent red, green, and blue components and recombining them through matrix addition to reconstruct the original image \cite{mathlens2026}. Figure \ref{3}(a) represents the $4\times4$ color image with their corresponding pixel intensity values.\

Furthermore, multiplying each of the RGB components by a scalar $k$ modifies the brightness of the image, as discussed in Section \ref{sec:scaling}).\ The corresponding visual effect for $k=1.55$ is illustrated in Figure \ref{3}(b).\ Importantly, the same scaling factor is applied to all three components. For example, consider a pixel represented by $\bvec{v} = [160, 80, 40]^T$. Scaling by $k=1.5$ yields
\begin{equation}
\bvec{v}' = \begin{bmatrix} 1.5 \times 160 \\ 1.5 \times 80 \\ 1.5 \times 40 \end{bmatrix} = \begin{bmatrix} 240 \\ 120 \\ 60 \end{bmatrix}
\end{equation}
where the ratio between the RGB components remain unchanged, i.e., 
\begin{equation}
\frac{R}{G} = \frac{160}{80} = 2 = \frac{240}{120}, \qquad \frac{G}{B} = \frac{80}{40} = 2 = \frac{120}{60}
\end{equation}
It is to be noted that the RGB ratios remain unchanged only when no component exceeds the permitted intensity range of $0-255$.\ If scaling causes a component to exceed 255, clipping is required, as discussed in Section \ref{sec:scaling}.\ \textit{MathLens} provides an interactive slider to vary the scaling factor ($k$) and observe the corresponding changes in image brightness \cite{mathlens2026}, as illustrated in Figure~\ref{3}(b).

\subsection{Two-Dimensional Spatial Linear Transformations}
So far, we have explored how mathematical operations can change the intensity and color of pixels. We now turn our attention to their spatial coordinates and investigate how matrices can be used to change the position, size, and orientation of an image.

Consider a point in an image represented by the coordinate vector $\mathbf{x} = [x, y]^T$, its new position after applying a linear transformation is given by
\begin{equation}
\bvec{x}' = \mathbf{A}\bvec{x} 
\end{equation}
Or equivalently,
\begin{equation}
 \begin{bmatrix} x' \\ y' \end{bmatrix} = \begin{bmatrix} a & b \\ c & d \end{bmatrix} \begin{bmatrix} x \\ y \end{bmatrix}
\end{equation}
where $\mathbf{A}$ is a $2\times2$ transformation matrix.\ By changing its elements, we can perform different geometric transformations, as illustrated below.

\subsubsection{Stretching and Compression (Scaling)}
\begin{figure}[t]
 \caption{Stretching along the $x$- and $y$-directions using transformation matrices with a scaling factor of $k=1.5$, along with the corresponding visual effects.}
\centering
    \includegraphics[width=1.01\columnwidth,keepaspectratio]{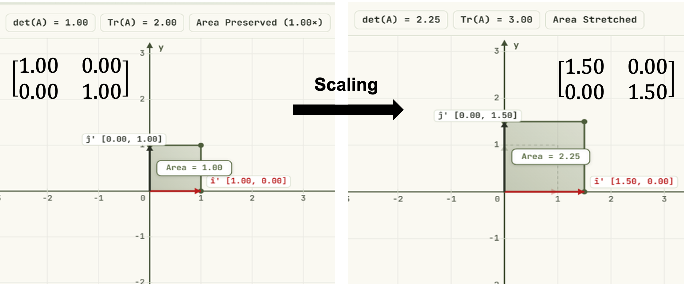}
    \label{4}
\end{figure}

Geometric scaling changes the size of an image along the horizontal and vertical directions \cite{lay2016}, by modifying the diagonal elements of the transformation matrix given by
\begin{equation}
\begin{bmatrix} x' \\ y' \end{bmatrix} = \begin{bmatrix} s_x & 0 \\ 0 & s_y \end{bmatrix} \begin{bmatrix} x \\ y \end{bmatrix}
\end{equation}
where $s_x$ and $s_y$ are the scaling factors along the \textit{x}- and \textit{y}-axes, respectively. For example, $\begin{bmatrix} 1.5 & 0 \\ 0 & 1.5 \end{bmatrix}$ increases the size of an image uniformly in both directions, whereas $\begin{bmatrix} 1.5 & 0 \\ 0 & 1 \end{bmatrix}$ stretches the image strictly along the horizontal axis, as shown in Figure \ref{4}.\ 
In \textit{MathLens}, users can change the scaling factor and observe how the diagonal elements of the transformation matrix control the dimension of the image \cite{mathlens2026}.

\subsubsection{Planar Rotation}
An image can be rotated by an angle $\theta$ about the origin using the rotation matrix given by
\begin{equation}
\begin{bmatrix} x' \\ y' \end{bmatrix} = \begin{bmatrix} \cos\theta & -\sin\theta \\ \sin\theta & \cos\theta \end{bmatrix} \begin{bmatrix} x \\ y \end{bmatrix}
\end{equation}
For example, a counterclockwise rotation through 45$^\circ$ transform the original identity matrix to
\begin{equation}
\mathbf{R}(45^\circ) = \begin{bmatrix} 0.71 & -0.71 \\ 0.71 & 0.71 \end{bmatrix}
\end{equation}
and through 90$^\circ$ transform it as
\begin{equation}
\mathbf{R}(90^\circ) = \begin{bmatrix} 0 & -1 \\ 1 & 0 \end{bmatrix}
\end{equation}

\begin{figure}[t]
 \caption{Rotation by an angle of 45$^\circ$ using a transformation matrix and its corresponding visual effect.}
\centering
    \includegraphics[width=1.01\columnwidth,keepaspectratio]{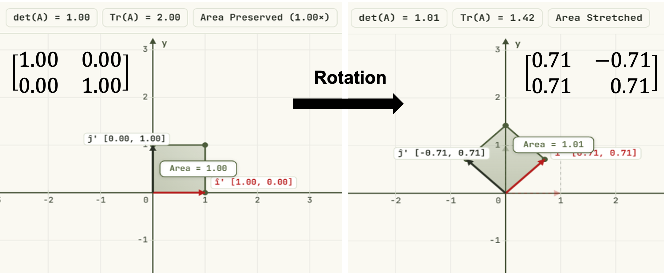}
    \label{5}
\end{figure}
The corresponding changes from identity matrix to 45$^\circ$ is shown in Figure \ref{5}. 
\textit{MathLens} provides an interactive angle slider ($0^\circ \to 30^\circ \to 45^\circ \to 90^\circ \to 180^\circ$) to observe the corresponding changes in the image \cite{mathlens2026}.

\subsubsection{Mirror Reflection}
\label{Mirror}
Reflection produces a mirror image about a chosen coordinate axis \cite{lay2016}. Reflection about the horizontal ($x$) and the vertical ($y$) axis is given by,
\begin{equation}
\mathbf{M}_{\text{horiz}} = \begin{bmatrix} 1 & 0 \\ 0 & -1 \end{bmatrix},\qquad 
\mathbf{M}_{\text{vert}} = \begin{bmatrix} -1 & 0 \\ 0 & 1 \end{bmatrix}
\end{equation}

\begin{figure}[t]
 \caption{Reflection across the vertical $y$-axis using a transformation matrix and its corresponding visual effect.}
\centering
    \includegraphics[width=\columnwidth,keepaspectratio]{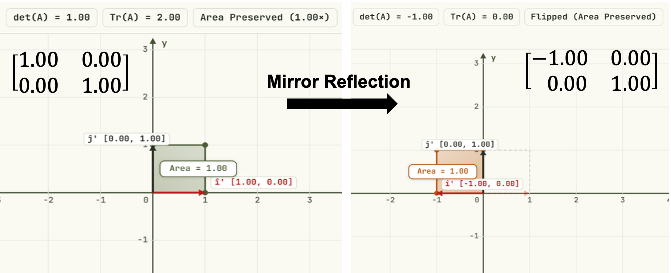}
    \label{6}
\end{figure}

It is to be noted that the negative sign reverses the corresponding coordinate, and the corresponding vertical shift for reflection about the $y$-axis is shown in Figure \ref{6}.
\textit{MathLens} provides a toggle between the original image, the horizontal flip, and the vertical flip, by changing the negative sign on the diagonal \cite{mathlens2026}.

\subsubsection{Coordinate Shearing}
\begin{figure}[t]
 \caption{Horizontal shearing with a shear factor of $k = 0.5$ using a transformation matrix and its corresponding visual effect.}
\centering
    \includegraphics[width=1.01\columnwidth,keepaspectratio]{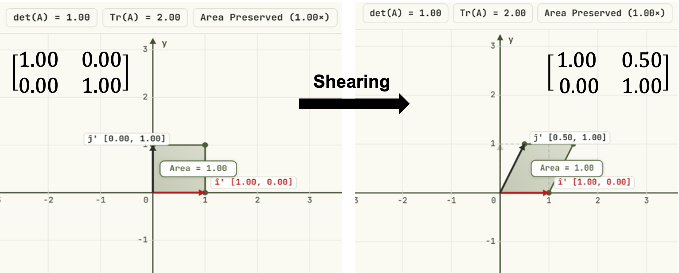}
    \label{7}
\end{figure}

Shearing changes the shape of an image by shifting one coordinate proportionally to the other \cite{lay2016}. Horizontal and vertical shearing are represented by
\begin{equation}
\mathbf{H}_x(k) = \begin{bmatrix} 1 & k \\ 0 & 1 \end{bmatrix}, \qquad \mathbf{H}_y(k) = \begin{bmatrix} 1 & 0 \\ k & 1 \end{bmatrix}
\end{equation}
By varying $k = 0 \to 0.2 \to 0.5 \to 1.0$, it can be seen that how a rectangular image gradually becomes slanted, demonstrating the role of off-diagonal elements of a transformation matrix. Figure \ref{7} represents the corresponding change for $k = 0.5$.

\subsubsection{Determinant: Stretch, Shrink, Flip, or Collapse?}
For a transformation matrix $\mathbf{A} = \begin{bmatrix} a & b \\ c & d \end{bmatrix}$, the determinant is given by
\begin{equation}
\det(\mathbf{A}) = ad - bc
\end{equation}

\begin{figure}[t]
 \caption{Collapse of a 2D image into a line under a transformation matrix with zero determinant, along with its corresponding visual effect.}
\centering
    \includegraphics[width=\columnwidth,keepaspectratio]{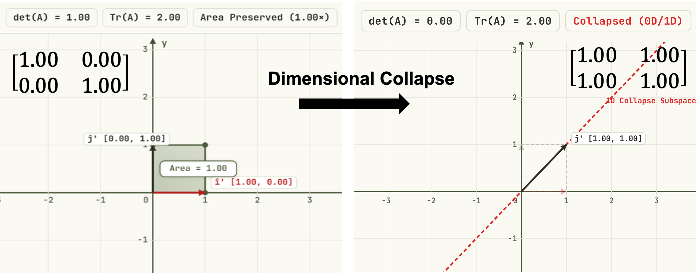}
    \label{8}
\end{figure}

\textit{MathLens} provides an interactive way to understand how the determinant of a transformation matrix relates to changes in area and orientation \cite{mathlens2026}, where
\begin{itemize}
    \item $|\det(\mathbf{A})| > 1$ indicates an increase in area (geometric expansion / stretch), as shown in Figure \ref{4} and \ref{5}.
    \item $0 < |\det(\mathbf{A})| < 1$ indicates decrease in area (geometric compression / shrink).
  \item $\det(\mathbf{A}) = 0$ indicates dimensional collapse, where a 2D image is reduced to a line or a point, as shown in Figure \ref{8}. The transformation loses information and cannot be inverted.
\item $\det(\mathbf{A}) < 0$ indicates a reversal of orientation, similar to a mirror reflection as shown in Figure \ref{6}.
\end{itemize}

\subsection{Matrix Addition and Subtraction}
\subsubsection{Matrix Addition and Image Blending} 
\begin{figure}[t]
 \caption{(a) Image blending with a blending factor of $\alpha = 0.52$ and (b) defect detection using direct matrix addition of $4 \times 4$ grayscale images, along with their corresponding matrix representations.}
\centering
\includegraphics[
        width=0.95\linewidth,
        height=0.75\textheight,
        keepaspectratio
    ]{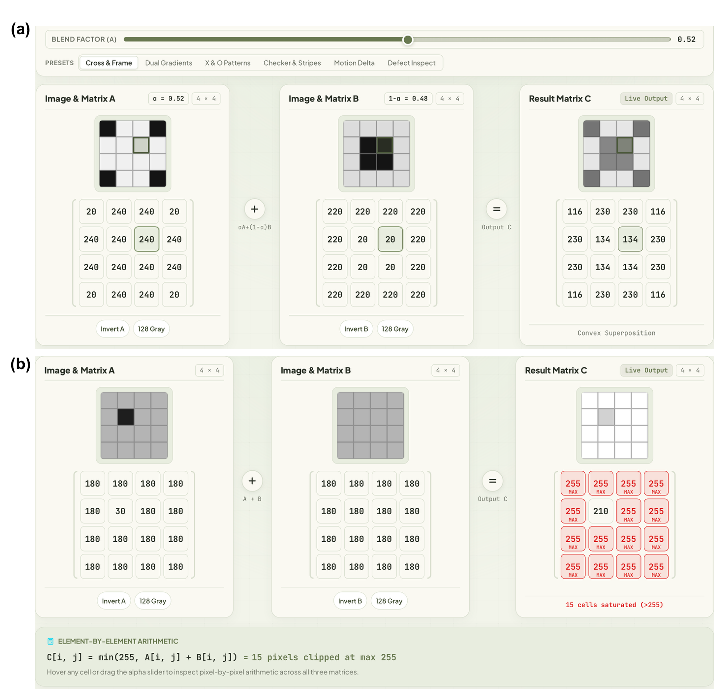}
     \label{9}
    \end{figure}

Matrix addition provides a simple way to combine two images of the same dimensions by adding their corresponding pixel values \cite{gonzalez2018image}.\ In Image blending, the contribution of each image can be controlled by introducing a weight factor ($\alpha$) given by
\begin{equation}
\mathbf{C} = \alpha\mathbf{A} + (1-\alpha)\mathbf{B}, \qquad 0 \le \alpha \le 1
\label{eq:image-blending}
\end{equation}

where $\mathbf{A}$ and $\mathbf{B}$ represent the original images, $\mathbf{C}$ is the resulting blended image, and $\alpha$ determines the relative contribution of each image. When $\alpha=0.5$, both images contribute equally, whereas $\alpha=0$ and $\alpha=1$ reproduce images $\mathbf{B}$ and $\mathbf{A}$, respectively. The image blending of two images with $\alpha=0.52$ is shown in Figure \ref{9}(a).
However, direct addition of two image matrices may produce pixel values exceeding the maximum permissible intensity of 255, requiring the resulting values to be clipped to this limit (Section \ref{sec:scaling}). This is illustrated in Figure \ref{9}(b), where direct matrix addition is employed to identify a defective element in an image.

\subsubsection{Matrix Subtraction: Background removal, Change detection, and Image Inversion}
\begin{figure}[t]
 \caption{(a) Image subtraction (b) background removal and change detection, and (c) generation of an X-ray-like photographic negative using matrix subtraction operation on $4 \times 4$ grayscale images, along with their corresponding matrix representations.}
\centering
    \includegraphics[
        width=0.95\linewidth,
        height=0.60\textheight,
        keepaspectratio
    ]{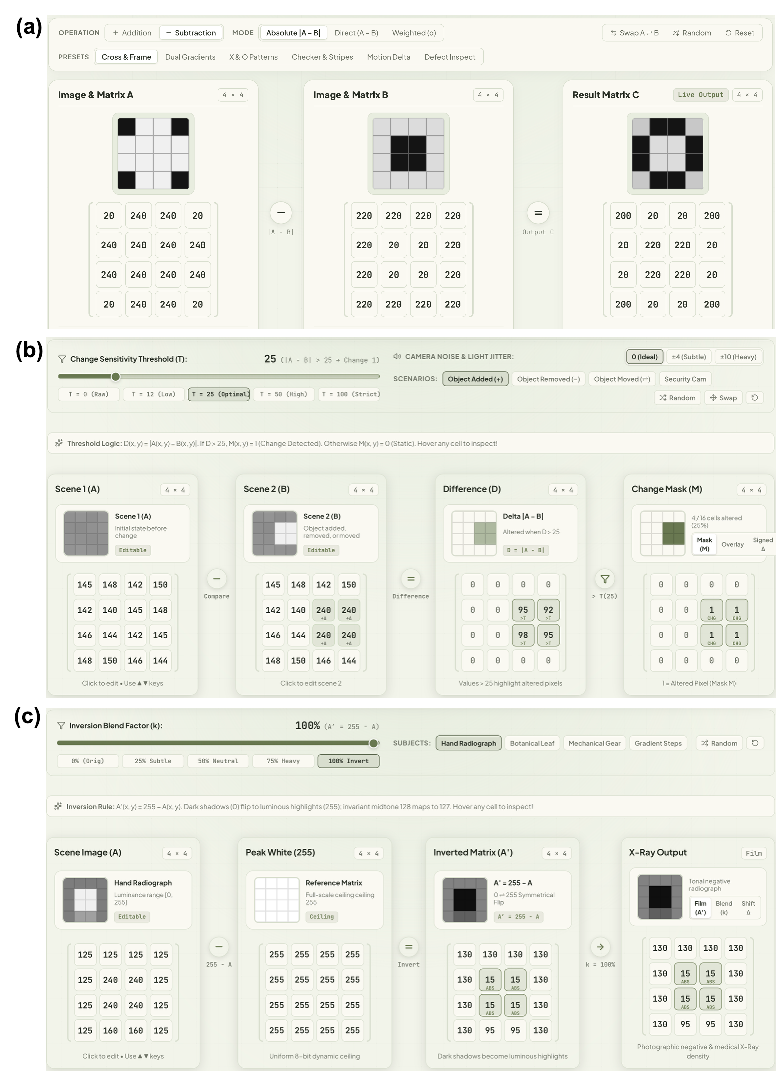}
     \label{10}
    \end{figure}

Matrix subtraction provides a simple way to identify the differences between two images \cite{gonzalez2018image}.\ Consider a background image represented by a matrix $\mathbf{B}$ and another image $\mathbf{I}$ of the same scene with some additional object. 
The difference matrix obtained by subtracting the corresponding pixel values  is given by
\begin{equation}
\mathbf{D} = |\mathbf{I} - \mathbf{B}|
\end{equation}
where the absolute value is taken element-wise. The corresponding visual effects are depicted in Figure \ref{10}(a).\ 

Ideally, the pixels belonging to the unchanged background have a difference of zero, whereas pixels corresponding to the newly introduced object have non-zero values. However, small variations in lighting and camera noise can also produce non-zero differences. To distinguish significant changes from these variations, a threshold $T$ can be introduced as
\begin{equation}
\mathbf{M}(x,y) = \begin{cases}
1, & \text{if } \mathbf{D}(x,y) > T \quad (\text{Foreground Object})\\
0, & \text{if } \mathbf{D}(x,y) \le T \quad (\text{Background Scene})
\end{cases}
\end{equation}

where M$(i,j)=1$ identifies pixels with significant changes, while M$(i,j)=0$ represents the unchanged background. This binary mask can be used to isolate the foreground object. The same principle can be extended to change detection or remove background by comparing two images of the same scene captured at different times, as illustrated in Figures \ref{10}(b). Subtracting their corresponding pixel values highlights regions where objects have appeared, disappeared, or moved.\ \textit{MathLens} also supports both direct and weighted image subtraction, where pixel values exceeding 255 are clipped to 255, while negative values are set to zero \cite{mathlens2026}.

Another interesting application of matrix subtraction is image inversion, which produces a photographic negative by reversing the pixel intensities. For an 8-bit grayscale image represented by matrix $\mathbf{A}$, the inverted image can be obtained as

\begin{equation}
\mathbf{A}' = 255 \mathbf{J} - \mathbf{A}
\end{equation}

where $\mathbf{J}$ is a matrix of ones having the same dimensions as $\mathbf{A}$. Thus, each pixel intensity transforms as
\begin{equation}
 a'_{ij} = 255 - a_{ij}
\end{equation}

As a result, black pixels ($0$) become white ($255$), white pixels become black, and intermediate grayscale values are similarly reversed. This produces a photographic negative or X-ray-like visual appearance, as shown in Figure \ref{10}(c).

\subsection{Cross-Image Matrix Multiplication}
\begin{figure}[t]
 \caption{Cross-image matrix multiplication of (a) two different image matrices and (b) an image matrix with the identity matrix, along with their corresponding visual effects.}
\centering
    \includegraphics[width=\columnwidth,keepaspectratio]{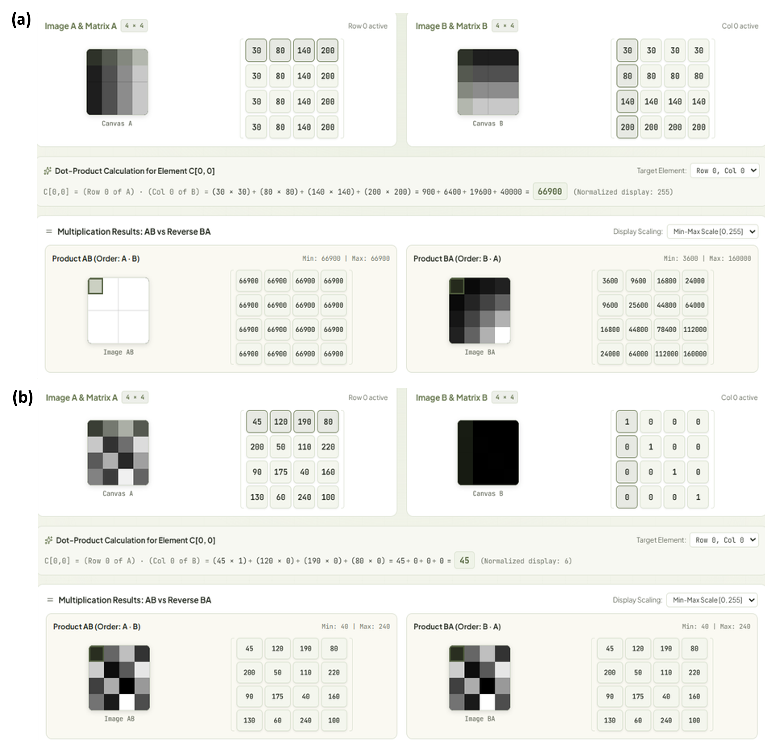}
     \label{11}
    \end{figure}

Unlike matrix addition and subtraction, which operate on corresponding pixel values, matrix multiplication combines the rows of one matrix with the columns of another. For example, consider two grayscale images $\mathbf{A}$ and $\mathbf{B}$ of order $4\times4$, as shown in Figure \ref{11}(a) and perform the standard matrix multiplication given by
\begin{equation}
\mathbf{C} = \mathbf{A}\mathbf{B}
\end{equation}
\noindent or, in element-wise form, 
\begin{equation}
c_{ij} = \sum_{k=1}^{4} a_{ik}b_{kj}
\end{equation}
where each output entry $c_{ij}$ is the dot product of the $i$-th row of $\mathbf{A}$ and the $j$-th column of $\mathbf{B}$.

It is to be noted that reversing the order of the matrix, $\mathbf{B}\mathbf{A}$ will give the difference result than the product $\mathbf{A}\mathbf{B}$, hence the different output image, as shown in Figure \ref{11}(a).
Since all rows of $\mathbf{A}$ and all columns of $\mathbf{B}$ are identical, every element of the resulting matrix has the same value. After normalization to the display range of $0–255$, the resulting image appears as a uniform gray square.\ However, reversing the multiplication order $\mathbf{B}\mathbf{A}$, the matrix elements vary from 3,600 to 160,000, producing a gradual change in pixel intensities across the image after normalization, as illustrated in Figure \ref{11}(a). Thus, the two products produce different numerical matrices and visual patterns, indicating the non-commutativity of the matrix multiplication, i.e., $\mathbf{A}\mathbf{B} \neq \mathbf{B}\mathbf{A}$.

Further, if $\mathbf{B}$ matrix is an identity matrix, then its product on reversing the matrix order would remain the same, i.e., $\mathbf{A}\mathbf{B} = \mathbf{B}\mathbf{A}$, as shown in Figure \ref{11}(b).

\subsection{Creating and Removing Shadows (Hadamard Product)}
\label{Hadamard}
\begin{figure}[t]
 \caption{(a) Hadamard product of \(4 \times 4\) grayscale image matrices, along with their corresponding matrix representations, and (b) its application in creating a shadow effect in a real-world image.}
\centering
    \includegraphics[width=\columnwidth,keepaspectratio]{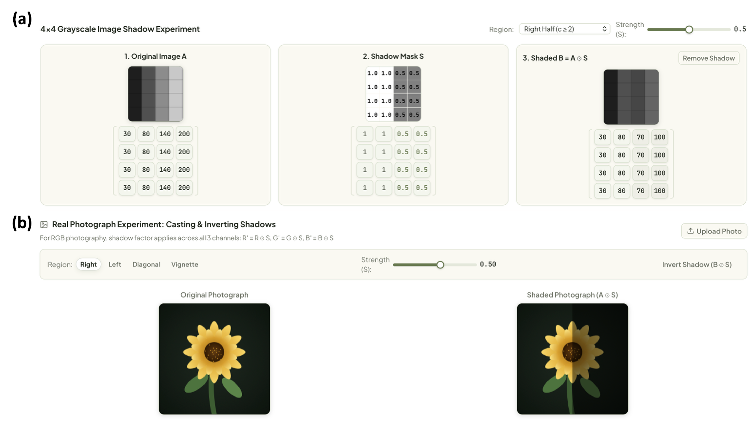}
     \label{12}
    \end{figure}

Matrix multiplication can also be extended to simulate the formation and removal of shadows in an image using the Hadamard product, denoted by $\odot$, which involves element-wise multiplication of two matrices \cite{horn1991topics}, as given by
\begin{equation}
\mathbf{B} = \mathbf{A} \odot \mathbf{S}\\ 
\end{equation}
\noindent or, in element-wise form,
 \begin{equation}
 b_{ij} = a_{ij} \cdot s_{ij}
\label{eq:hadamard-shadow}
\end{equation}
Here, $\mathbf{A}$ represents the original image, while $\mathbf{S} \in [0,1]^{M \times N}$ is a shadow attenuation matrix that controls the intensity of individual pixels. A value of $s_{ij}=1$ leaves the pixel intensity unchanged, $s_{ij}=0.5$ reduces it by 50$\%$, and $s_{ij}=0$ produces complete darkness. For example, to simulate a shadow covering the right half of an image, the attenuation factors are set to $s_{ij}=1$ for the illuminated region and $s_{ij}=0.5$ for the shadowed region.\ Figure~\ref{12}(a) illustrates the Hadamard product of two matrices through the original image matrix $\mathbf{A}$, the attenuation matrix $\mathbf{S}$, and the resulting shadowed image $\mathbf{A}\odot\mathbf{S}$.

Furthermore, the original image can be recovered from the shadowed image through element-wise division, provided that the attenuation factors are non-zero ($s_{ij}>0$), and is given by
\begin{equation}
\hat{\mathbf{A}} = \mathbf{B} \oslash \mathbf{S}
\end{equation}
\noindent or, in element-wise form,
\begin{equation}
\hat{a}_{ij} = \frac{b_{ij}}{s_{ij}}
\end{equation}
where $\hat{\mathbf{A}}$ represents the reconstructed original image and $\oslash$ denotes the element-wise division.\ For example, consider a pixel whose original intensity is reduced to $b_{ij}=100$ by a shadow attenuation factor of $s_{ij}=0.5$, its original intensity can be restored as
\begin{equation}
\hat{a}_{ij} = \frac{100}{0.5} = 200
\end{equation}
\textit{MathLens} provide both the formation and removal of shadows through simple matrix operations by adjusting the attenuation factors \cite{mathlens2026}.\ However, exact recovery is possible only when the attenuation factors are known and non-zero, and no information is lost due to rounding or quantization of pixel intensities.

\subsubsection{Extension to Real Color Photographs}
This concept can also be extended to real color photographs, where the same shadow attenuation matrix is independently applied to each of the three RGB channels given by
\begin{equation}
\mathbf{R}' = \mathbf{R} \odot \mathbf{S}, \quad \mathbf{G}' = \mathbf{G} \odot \mathbf{S}, \quad \mathbf{B}' = \mathbf{B} \odot \mathbf{S}
\end{equation}
Through the interactive interface in \textit{MathLens}, we can select a region of the image, vary the shadow intensity using a slider, and observe the corresponding changes in real time, as shown in Figure \ref{12}(b). Further, the ``Remove Shadow'' option allows to explore the reverse operation and recover the original image \cite{mathlens2026}, subject to the limitations discussed above (Section \ref{Hadamard}).

\subsection{Matrix Inverse: Undoing Transformations}
\label{Inverse}
A square matrix $\mathbf{A}$ is invertible if its determinant is non-zero, i.e., $\det(\mathbf{A})\neq 0$. The inverse matrix $\mathbf{A}^{-1}$ allows us to reverse a geometrical transformation and recover the original coordinates given by

\[
\bvec{x}'=\mathbf{A}\bvec{x}
\iff
\bvec{x}=\mathbf{A}^{-1}\bvec{x}',
\qquad \text{and} \qquad
\mathbf{A}^{-1}\mathbf{A}=\mathbf{I}
\]

where $\mathbf{I}$ is the identity matrix.

\textit{MathLens} demonstrates this concept through the ``Undo Transformation Using Inverse'' feature, which applies the inverse transformation to restore the original coordinates \cite{mathlens2026}. For example, consider a scaling transformation that doubles the horizontal coordinate while leaving the vertical coordinate unchanged, i.e.,
\[
\mathbf{A}=
\begin{bmatrix}
2 & 0\\
0 & 1
\end{bmatrix}
\implies
\mathbf{A}^{-1}=
\begin{bmatrix}
0.5 & 0\\
0 & 1
\end{bmatrix}, \qquad \text{and} \qquad
\mathbf{A}^{-1}\mathbf{A} = \begin{bmatrix} 1 & 0 \\ 0 & 1 \end{bmatrix} = \mathbf{I}
\]

\begin{figure}[t]
 \caption{Forward scaling transformation of coordinates in the $xy$-plane using a transformation matrix, followed by the inverse transformation that restores the original geometry, along with their corresponding matrix representation and visual effects.}
\centering
    \includegraphics[width=1.01\columnwidth,keepaspectratio]{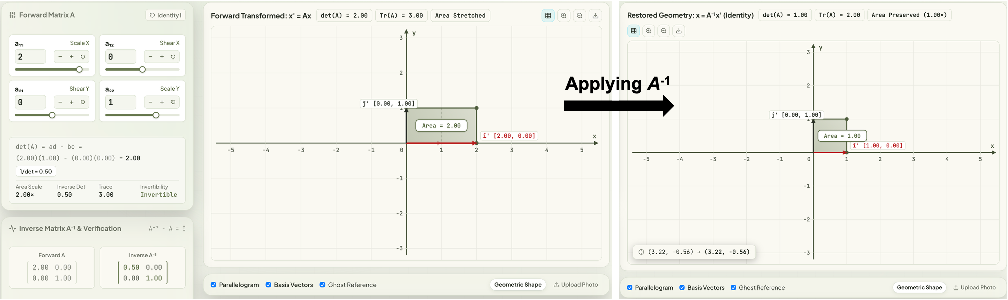}
     \label{13}
    \end{figure}
The corresponding visual effect is shown in Figure \ref{13}. In \textit{MathLens}, these transformations include scaling, rotation, shearing, and reflection \cite{mathlens2026}.\ It is to be noted that exact recovery of the original image may not always be possible due to pixel rounding or cropping during the transformation.

\subsection{Zoom In and Zoom Out}
The same principle can also be extended to zooming in and out of images. Zooming can be represented by a scaling matrix, while its inverse reverses the scaling and restores the original coordinates. For a uniform scaling factor $k>0$, the scaling matrix $\mathbf{S}$ and its inverse $\mathbf{S}^{-1}$ are given by

\[
\mathbf{S}=
\begin{bmatrix}
k&0\\
0&k
\end{bmatrix},
\qquad
\mathbf{S}^{-1}=
\begin{bmatrix}
\frac{1}{k}&0\\
0&\frac{1}{k}
\end{bmatrix},
\qquad \text{and} \qquad
\mathbf{S}^{-1}\mathbf{S}=\mathbf{I}
\]

Here, $k>1$ corresponds to zooming in, while $0<k<1$ corresponds to zooming out. \textit{MathLens} illustrates both operations and their reversal through inverse transformations given by

\[
\begin{aligned}
\text{Original}
&\xrightarrow{\text{Zoom In }2\times}
\text{Scaled}
\xrightarrow{\mathbf{S}^{-1}}
\text{Original}\\
\text{Original}
&\xrightarrow{\text{Zoom Out }0.25\times}
\text{Scaled}
\xrightarrow{\mathbf{S}^{-1}}
\text{Original}
\end{aligned}
\]

\begin{figure}[t]
 \caption{(a) Zooming in with a scaling factor of $k=2$ and (b) zooming out with a scaling factor of $k=0.25$, applied to an identity matrix, along with their corresponding matrix representations and visual effects.}
\centering
   \includegraphics[
        width=0.95\linewidth,
        height=0.75\textheight,
        keepaspectratio
    ]{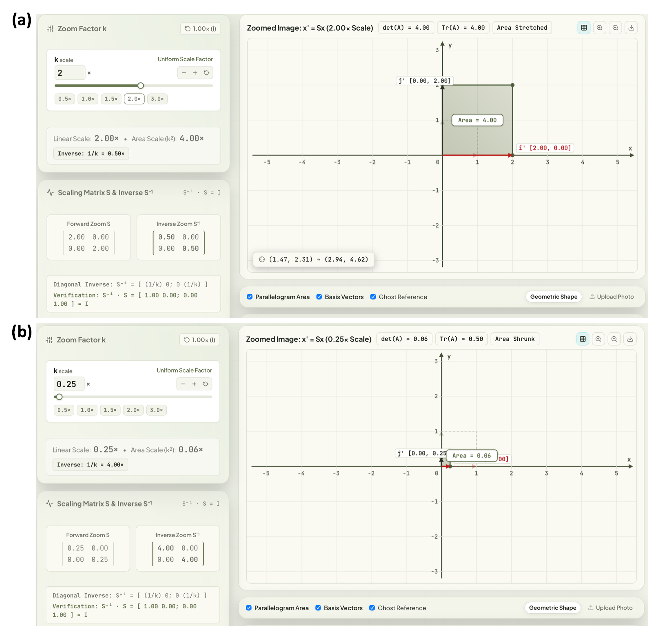}
     \label{14}
    \end{figure}

For example, during a $2\times$ zoom and $0.25\times$ zoom, the basis vectors $(1,0)$ and $(0,1)$ are transformed into [$(2,0)$, $(0,2)$] and [$(0.25,0)$, $(0,0.25)$], respectively. The determinants $\det(\mathbf{S})=4$ and $\det(\mathbf{S})=0.06$ indicate that the area is scaled by factors of 4 and 0.06, respectively, corresponding to expansion and a contraction, as illustrated in Figure \ref{14}(a) and \ref{14}(b).\ Applying the inverse scaling matrix restores the original basis vectors and unit grid, as discussed in Section \ref{Inverse}. 

\section{From Observation to Exploration: My Image, My Matrix}
\begin{figure}[t]
 \caption{(a) Scaling of image pixel intensities, showing the corresponding matrix representation, pixel locations, and RGB component values, and (b) comparison of the original and transformed matrices at the corresponding pixel locations.}
\centering
  \includegraphics[
        width=0.95\linewidth,
        height=0.60\textheight,
        keepaspectratio
    ]{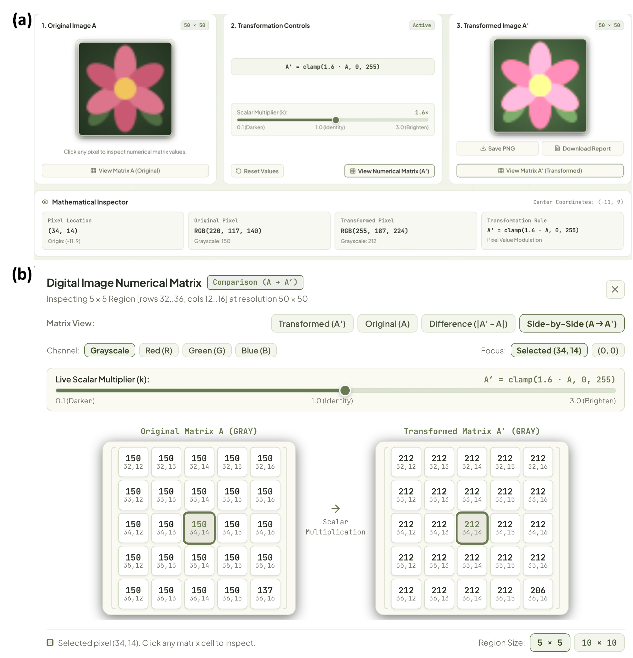}
     \label{15}
    \end{figure}

The mathematical concepts discussed in the preceding sections can be further explored through the ``My Image, My Matrix'' module of \textit{MathLens}, which allows us to upload images of our choice and examine their matrix representations and transformations at different resolutions \cite{mathlens2026}. We can also select a region of an image to inspect individual pixel values and observe how they change when different mathematical operations are applied, as illustrated in Figure \ref{15}. This extends the learning experience beyond predefined examples, allowing us to experiment with familiar images and explore the connection between mathematical operations, their corresponding matrix elements, and the resulting visual changes.

\section{Ray Optics: Planar Reflection and Successive Reflections}
\begin{figure}[t]
 \caption{(a) Representation of a light ray using a direction vector and (b) rotation of the ray using a transformation matrix, along with their corresponding mathematical and visual representations.}
\centering
    \includegraphics[width=\columnwidth,keepaspectratio]{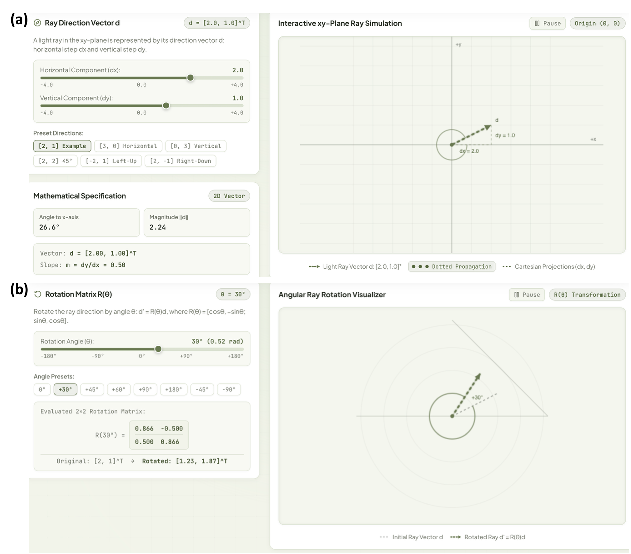}
     \label{16}
    \end{figure}

Beyond transforming digital images, the same matrix operations can also be used to describe physical phenomena, such as change in the direction of light rays, and reflections from single or multiple plane mirrors \cite{ghatak2025optics, pedrotti2017optics}. To illustrate this, consider a light ray propagating in the $xy$-plane. Its direction of propagation can be represented by a two-dimensional vector given by,
\begin{equation}
\mathbf{d} =
\begin{bmatrix}
d_x \\
d_y
\end{bmatrix}
\end{equation}
where $d{_x}$ and $d{_y}$ denote the horizontal and vertical components, respectively.
For example, the vector
\[
\mathbf{d} =
\begin{bmatrix}
2 \\
1
\end{bmatrix}
\]
represents a ray whose horizontal displacement is twice as its vertical displacement, as shown in Fig \ref{16}(a). The direction of the ray is therefore determined by the relative magnitudes of these two components. \textit{MathLens} provides the sliders to modify $d{_x}$ and $d{_y}$, and observe how the direction of the ray changes in the $xy$-plane \cite{mathlens2026}.

\subsection{Rotation of a Light Ray}
The change in the direction of a light ray can also be described mathematically using a rotation matrix. Consider a ray propagating in the $xy$-plane with direction vector $\mathbf{d}$. If this vector is rotated counterclockwise by an angle $\theta$, the new direction is given by
\begin{equation}
    \mathbf d'=\mathbf R(\theta)\mathbf d
\end{equation}
where
\begin{equation}
\mathbf{R(\theta)}=
\begin{bmatrix}
\cos\theta&-\sin\theta\\
\sin\theta&\cos\theta
\end{bmatrix}
\end{equation}
is the two-dimensional rotation matrix.\ 

For example, consider a ray initially traveling along the positive $x$-axis, represented by
\begin{equation}
\mathbf d=\begin{bmatrix}2\\1\end{bmatrix}
\end{equation}
a counterclockwise rotation of 30$^\circ$ transform its direction as
\begin{equation}
\begin{aligned}
\mathbf{d}' &= \mathbf{R}(30^\circ)\mathbf{d}\\
&=
\begin{bmatrix}
0.866 & -0.500\\
0.500 & 0.866
\end{bmatrix}
\begin{bmatrix}
2\\
1
\end{bmatrix}
=
\begin{bmatrix}
1.232\\
1.866
\end{bmatrix}
\end{aligned}
\end{equation}
The resulting change in the ray's direction is depicted in Figure \ref{16}(b).\ It is important to note that the rotation matrix describes the change in the ray's direction mathematically, whereas in physical ray optics, such changes arise from interactions such as reflection or refraction.

\textit{MathLens} provides an interactive slider to vary the rotation angle from -180$^\circ$ to 180$^\circ$, and observe the corresponding changes in the ray's direction and rotation matrix \cite{mathlens2026}. 

\subsection{Reflection of a Light Ray from a Plane Mirror}
\begin{figure}[t]
 \caption{(a) Planar Reflection from the horizontal mirror (b) Successive reflection from the horizontal followed by vertical mirror, along with their matrix and visual representations.}
\centering
    \includegraphics[width=\columnwidth,keepaspectratio]{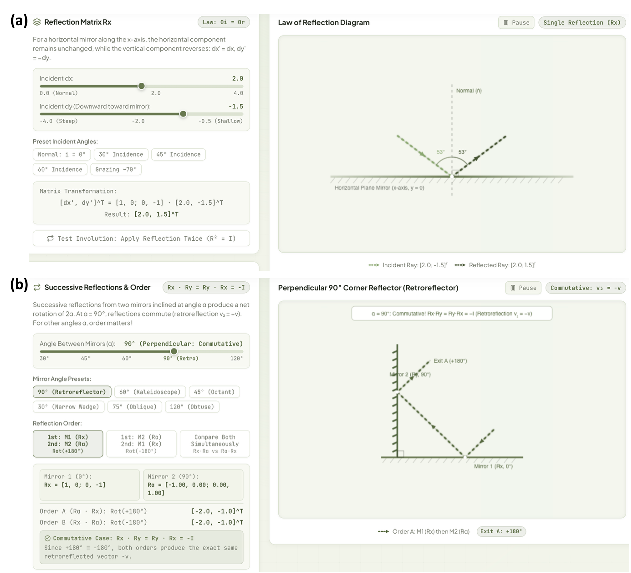}
     \label{17}
    \end{figure}

Consider a light ray incident on a horizontal plane mirror placed along the $x$-axis. Upon reflection, the component of the ray's direction parallel to the mirror remains unchanged, while the perpendicular component reverses sign. This change in direction can be represented mathematically using a reflection matrix given by
\begin{equation}
\mathbf{d}' = \mathbf{M}_x \mathbf{d},
\qquad \text{and} \qquad
\mathbf{M}_x =
\begin{bmatrix}
1 & 0 \\
0 & -1
\end{bmatrix}
\label{Rx}
\end{equation}
where $\mathbf{M}_x$ represents the reflection matrix for a horizontal mirror aligned along the $x$-axis (Section \ref{Mirror}), and $\mathbf{d}'$ represents the direction of the reflected ray.
For example, consider an incident ray with direction vector
\begin{equation}
  \mathbf{d}=
\begin{bmatrix}
2\\
-1.5
\end{bmatrix}  
\end{equation}
Upon reflection about the $x$-axis, the direction of the ray changes according to
\begin{equation}
\begin{aligned}
\mathbf{d}' &= \mathbf{M_{\textit{x}}}\mathbf{d}\\
&=
\begin{bmatrix}
1 & 0\\
0 & -1
\end{bmatrix}
\begin{bmatrix}
2\\
-1.5
\end{bmatrix}
=
\begin{bmatrix}
2\\
1.5
\end{bmatrix}
\end{aligned}
\end{equation}
The resulting change in the ray's direction is shown in Figure \ref{17}(a), where the angle of reflection remains equal to the angle of incidence, in accordance with the law of reflection. \textit{MathLens} provides an interface to modify the direction of the incident ray and different angles of incidence, and observe how matrix multiplication determines the direction of the reflected ray \cite{mathlens2026}.\ Further, it is to be noted that the reflection matrix in Equation \ref{Rx} corresponds to a horizontal plane mirror, whose orientation w.r.t. the $x$-axis is $0^\circ$. However, for a mirror inclined at an angle $\theta$ with respect to the positive $x$-axis, the reflection matrix $\mathbf{M}(\theta)$ can be expressed as
\begin{equation}
\mathbf{M}(\theta)=
\begin{bmatrix}
\cos 2\theta & \sin 2\theta\\
\sin 2\theta & -\cos 2\theta
\end{bmatrix}.
\label{Rtheta}
\end{equation}
The factor $2\theta$ arises because reflection about
a mirror inclined at an angle $\theta$ to the $x$-axis
can be represented by rotating the coordinate system
to align with the mirror, reflecting about the
$x$-axis, and rotating back to the original
coordinate system. Mathematically, this can be
expressed as
\begin{equation}
\mathbf{M}(\theta)
=
\mathbf{R}(\theta)\mathbf{M}_x\mathbf{R}(-\theta)
\end{equation}
where $\mathbf{R}(\theta)$ represents the rotation
matrix and $\mathbf{M}_x$ denotes reflection about
the $x$-axis.

For example, when the mirror is inclined at $30^\circ$,
the corresponding reflection matrix becomes

\begin{equation}
\mathbf{M}(30^\circ)=
\begin{bmatrix}
\frac{1}{2} & \frac{\sqrt{3}}{2}\\
\frac{\sqrt{3}}{2} & -\frac{1}{2}
\end{bmatrix}
\end{equation}

For the same incident ray with direction vector
$\mathbf{d}=[2,-1.5]^T$, the reflected direction is

\begin{equation}
\begin{aligned}
\mathbf{d}' &= \mathbf{M}(30^\circ)\mathbf{d}\\
&=
\begin{bmatrix}
\frac{1}{2} & \frac{\sqrt{3}}{2}\\
\frac{\sqrt{3}}{2} & -\frac{1}{2}
\end{bmatrix}
\begin{bmatrix}
2\\
-1.5
\end{bmatrix}\\
&=
\begin{bmatrix}
1-\frac{3\sqrt{3}}{4}\\
\sqrt{3}+\frac{3}{4}
\end{bmatrix}
\approx
\begin{bmatrix}
-0.299\\
2.482
\end{bmatrix}
\end{aligned}
\end{equation}

Thus, the direction of the reflected ray depends on
both the incident direction and the orientation of
the mirror. \textit{MathLens} allows users to explore this
dependence interactively by varying the mirror angle
and observing the corresponding changes in the
reflected direction \cite{mathlens2026}.

\subsection{Two Successive Reflections of a Light Ray}
What happens when a light ray undergoes two successive reflections? Can we predict its final direction using matrix multiplication? To explore this, consider two plane mirrors placed perpendicular to each other, one horizontal and the other vertical, with one mirror along the $x$-axis and the other along the $y$-axis. A light ray first reflects from the horizontal
mirror and then from the vertical mirror, as illustrated in
Figure \ref{17}(b).

\subsubsection{First Reflection: Horizontal Mirror}
Consider an incident ray represented by the direction vector
\[
\mathbf{d}=
\begin{bmatrix}
2\\
-1
\end{bmatrix}
\]
Reflection from the horizontal mirror along the $x$-axis is represented by
\[
\mathbf{M}_x=
\begin{bmatrix}
1&0\\
0&-1
\end{bmatrix}
\]
Applying this matrix to the incident direction vector gives
\[
\mathbf{d}_1=\mathbf{M}_x\mathbf{d}
=
\begin{bmatrix}
1&0\\
0&-1
\end{bmatrix}
\begin{bmatrix}
2\\
-1
\end{bmatrix}
=
\begin{bmatrix}
2\\
1
\end{bmatrix}
\]

Thus, the horizontal component remains unchanged, while the vertical component reverses its sign.

\subsubsection{Second Reflection: Vertical Mirror}
The reflected ray now encounters a vertical mirror along the $y$-axis, represented by the reflection matrix
\[
\mathbf{M}_y=
\begin{bmatrix}
-1&0\\
0&1
\end{bmatrix}
\]

Applying this matrix to the direction vector $\mathbf{d}_1$ gives

\[
\mathbf{d}_2=\mathbf{M}_y\mathbf{d}_1
=
\begin{bmatrix}
-1&0\\
0&1
\end{bmatrix}
\begin{bmatrix}
2\\
1
\end{bmatrix}
=
\begin{bmatrix}
-2\\
1
\end{bmatrix}
\]

Interestingly, the final direction is exactly opposite to the initial direction, i.e., 
\[
\mathbf{d}_2=-\mathbf{d}
\]
Thus, two successive reflections from perpendicular mirrors reverse the direction of the incident ray, provided that the ray encounters both reflecting surfaces, as shown in Figure \ref{17}(b).

\subsubsection{Does the Order of Reflections Matter?}
This leads to another interesting question: What happens if we reverse the order of the two reflections? 
When reflection from the horizontal mirror is followed by reflection from the vertical mirror, the combined transformation is
\[
\mathbf{M}_y\mathbf{M}_x=
\begin{bmatrix}
-1&0\\
0&-1
\end{bmatrix}
\]
Upon reversing the order, we get
\[
\mathbf{M}_x\mathbf{M}_y=
\begin{bmatrix}
-1&0\\
0&-1
\end{bmatrix}
\]

Therefore,
\[
\mathbf{M}_y\mathbf{M}_x=\mathbf{M}_x\mathbf{M}_y=-\mathbf{I}
\]

This shows that for reflections about two perpendicular
axes, the order of matrix multiplication does not change
the final direction. However, this property does not
generally hold for reflections about two mirrors inclined
at arbitrary angles. For example, consider two plane
mirrors, one along the $x$-axis and the other inclined
at $30^\circ$ to the $x$-axis. Their corresponding
reflection matrices are

\begin{equation}
\mathbf{M}_x =
\begin{bmatrix}
1 & 0\\
0 & -1
\end{bmatrix},
\qquad
\mathbf{M}_{30^\circ} =
\begin{bmatrix}
\cos 60^\circ & \sin 60^\circ\\
\sin 60^\circ & -\cos 60^\circ
\end{bmatrix}
=
\begin{bmatrix}
\frac{1}{2} & \frac{\sqrt{3}}{2}\\
\frac{\sqrt{3}}{2} & -\frac{1}{2}
\end{bmatrix}.
\end{equation}

For an initial direction vector
$\mathbf{d}=[2,-1]^T$, successive reflections first
about the $x$-axis and then about the inclined mirror
give

\begin{equation}
\begin{aligned}
\mathbf{M}_{30^\circ}\mathbf{M}_x\mathbf{d}
&=
\begin{bmatrix}
\frac{1}{2} & \frac{\sqrt{3}}{2}\\
\frac{\sqrt{3}}{2} & -\frac{1}{2}
\end{bmatrix}
\begin{bmatrix}
2\\
1
\end{bmatrix}\\
&=
\begin{bmatrix}
1+\frac{\sqrt{3}}{2}\\
\sqrt{3}-\frac{1}{2}
\end{bmatrix}\\
&\approx
\begin{bmatrix}
1.866\\
1.232
\end{bmatrix}
\end{aligned}
\end{equation}

However, reversing the order of reflections gives

\begin{equation}
\begin{aligned}
\mathbf{M}_x\mathbf{M}_{30^\circ}\mathbf{d}
&=
\begin{bmatrix}
1 & 0\\
0 & -1
\end{bmatrix}
\begin{bmatrix}
1-\frac{\sqrt{3}}{2}\\
\sqrt{3}+\frac{1}{2}
\end{bmatrix}\\
&=
\begin{bmatrix}
1-\frac{\sqrt{3}}{2}\\
-\sqrt{3}-\frac{1}{2}
\end{bmatrix}\\
&\approx
\begin{bmatrix}
0.134\\
-2.232
\end{bmatrix}
\end{aligned}
\end{equation}

Thus, the two orders of reflection produce different
final directions, illustrating that matrix multiplication
is generally non-commutative. This difference can be
explored visually through the \textit{MathLens} application
\cite{mathlens2026}.

Overall, the ray optics module in \textit{MathLens} illustrates
how matrix operations extend beyond digital image
transformations to describe the reflection of light
rays, allowing learners to connect abstract mathematical
concepts with physical phenomena through interactive
visualizations.

\section{Conclusion}
 \textit{MathLens: Interactive Mathematics} \& \textit{Linear Algebra Lab},
provides an interactive and visual approach to understanding
fundamental mathematical concepts by connecting matrix operations
and transformations with familiar digital images and physical phenomena. Through interactive sliders and real-time visualizations, learners can observe how changes in mathematical parameters influence image properties, geometrical transformations, and the direction of light rays. Further, the My Image, My Matrix module allows learners to experiment with their own images, extending the learning experience beyond predefined examples. By bridging abstract mathematical concepts with their visual interpretations and practical applications, \textit{MathLens} offers a supplementary educational tool for introductory courses in mathematics, physics, engineering, and computer science. Overall, this approach aims to make linear algebra more accessible, intuitive, and engaging while encouraging learners to explore its relevance to everyday technologies and physical phenomena. 

 Building on the fundamental concepts introduced in this article, future series articles are planned to explore more advanced topics in linear algebra, including eigenvalues and eigenvectors, matrix rank, norm, systems of linear equations, principle component analysis (PCA) and matrix decompositions such as QR decomposition and singular value decomposition (SVD), with an emphasis on their geometric interpretations and practical applications.

\section*{Application Availability}
The \textit{MathLens} educational application developed as part
of this work is available at \url{https://mathlens-kappa.vercel.app/?step=17}. The application allows users to interactively explore
matrix operations, geometrical transformations, and
elementary ray optics through visualizations and
hands-on activities.

\section*{Acknowledgement}
P. Wadhwa gratefully acknowledge Prof. P. K. Ahluwalia for his valuable suggestions regarding the inclusion of the My image, My matrix and ray optics modules in the \textit{MathLens} application.  



\bibliographystyle{unsrt}
\bibliography{References}

\end{document}